\documentclass{JHEP3}
\usepackage{amsmath}
\usepackage{cite}
\usepackage{epsfig}
\usepackage{mathdots}

\newcommand{\cO} {{\cal O}}

\newcommand{\la} {{\lambda}}
\newcommand{\La} {{\Lambda}}
\newcommand{\al} {{\alpha}}

\newcommand{\nodag} {{\phantom{\dagger}}}

\newcommand{\ket}[1] {\left|#1\right>}
\newcommand{\braket}[2] {\left<#1\vphantom{#2}\right|
                         \left.\!\vphantom{#1}{#2}\right>}
\newcommand{\vev}[1] {\left<#1\right>}

\newcommand{\cB} {{\cal B}}

\newcommand{\cL} {{\cal L}}

\title{Marginal deformations in string field theory}

\author{Ehud Fuchs, Michael Kroyter\\
Max-Planck-Institut f\"ur Gravitationsphysik\\
Albert-Einstein-Institut\\
14476 Golm, Germany\\
\email{udif@aei.mpg.de, mikroyt@aei.mpg.de}
}
\author{Robertus Potting\\
CENTRA, Departamento de F\'\i sica\\
Faculdade de Ci\^encias e Tecnologia, Universidade do Algarve\\
8005-139 Faro, Portugal\\
\email{rpotting@ualg.pt}
}

\abstract{We describe a method for obtaining analytic solutions
corresponding to exact marginal deformations in open bosonic string
field theory.
For the photon marginal deformation we have an explicit analytic
solution to all orders.
Our construction is based on a pure gauge solution where the gauge
field is not in the Hilbert space.
We show that the solution itself is nevertheless perfectly regular.
We study its gauge transformations and calculate
some coefficients explicitly.
Finally, we discuss how our method can be implemented for other
marginal deformations.}

\preprint{{\tt arXiv:0704.2222}\\AEI-2007-023}

\begin{document}

\section{Introduction}

The recent analytic construction by Schnabl~\cite{Schnabl:2005gv} (see
also~\cite{Okawa:2006vm,Fuchs:2006hw,Fuchs:2006an,Rastelli:2006ap,Ellwood:2006ba,Fuji:2006me,Fuchs:2006gs,Imbimbo:2006tz,Erler:2006hw,Erler:2006ww})
of an exact solution in classical open string field theory
(OSFT)~\cite{Witten:1986cc}
corresponding to the tachyon vacuum has given a renewed impetus in
using OSFT as a tool in the analysis of open string vacua. In
particular, it allowed for a proof of the first two of Sen's
conjectures~\cite{Sen:1999mh,Sen:1999xm}.  A class of related
solutions based on general projectors has been developed
in~\cite{Okawa:2006sn}.

This advance raised the hope that other solutions can also be found,
such as lump solutions
and marginal deformations. That marginal deformations can be described
within the framework of string field theory was shown by Sen
\cite{Sen:1990hh,Sen:1990na,Sen:1992pw}.
There, it was shown that boundary marginal deformations can be described
within OSFT, while bulk marginal deformations can be described using a
non-polynomial closed string field
theory, such as~\cite{Zwiebach:1992ie}.
These solutions
were first investigated using level truncation in the
Siegel gauge~\cite{Sen:2000hx}.
Other studies of these solutions appeared in
\cite{Zwiebach:2000dk,Iqbal:2000qg,Takahashi:2002ez,Kluson:2002hr,Kluson:2003xu,
Katsumata:2004cc,Sen:2004cq,Yang:2005iu,Kishimoto:2005bs}.

More recently, a recursive procedure
has been developed by using the techniques employed in Schnabl's solution 
(in particular, the $\mathcal{B}_0$ gauge), yielding exactly marginal
deformations order by order in a parameter $\lambda$ parameterizing
the exactly flat direction~\cite{Kiermaier:2007ba,Schnabl:2007az}.
This approach was generalized~\cite{Erler:2007rh,Okawa:2007ri}
to describe also the first analytical solutions of superstring field
theory~\cite{Berkovits:1995ab}.

The approach of~\cite{Kiermaier:2007ba,Schnabl:2007az} gives an explicit
solution for marginal deformation generated by current operators
which have a regular OPE with themselves.
For the more interesting case, such as the photon marginal deformation,
where
the OPE of the vertex operator $V(z)$ defining the marginal
deformation with itself is $V(0)V(z)\sim 1/z^2$, divergences arise as
the separations of the boundary insertions used in constructing the
solution go to zero. This makes it necessary to add counter terms
in order to cancel these divergences.
However, the form of these counter terms is known only up to the third
order and it is not a priori clear that counter terms for higher orders
exist.

In this work we propose an alternative approach toward the analytical
construction of exactly marginal deformations. It is based on
solutions that are formally pure gauge, but nonetheless nontrivial as
the gauge parameter we employ is not in the physical Hilbert space.
As we will show, this method allows us to obtain an explicitly defined
solution, perturbative in the above-mentioned parameter $\lambda$.  As
the insertions of the vertex operator in our approach remain at a
finite distance, the divergences encountered
in~\cite{Kiermaier:2007ba,Schnabl:2007az} do not arise. However, it
turns out that the solutions we obtain at first instance are
singular in the sense that there is a non-normalizable dependence on
the center of mass coordinate $x_0$. Only with a carefully chosen set
of counter terms
is it possible to regularize this unwanted dependence
on $x_0$, such that the solutions are in fact independent of the
center of mass coordinate.

Our approach starts off with the pure gauge solutions for string
field theory of the form~\cite{Okawa:2006vm,Ellwood:2006ba}
\begin{equation}
\label{eq:Psi}
\Psi = (1-\la\phi)Q\frac{1}{1-\la\phi}
    = Q\phi\frac{\la}{1-\la\phi}\,,
\end{equation}
which have the structure of a pure gauge solution generated by the
gauge field
\begin{equation}
\Lambda = -\log(1-\lambda\phi)\,.
\end{equation}
This is the case, because the finite gauge transformation in string field
theory takes the form
\begin{equation}
\label{generalGauge}
\Psi\rightarrow e^{-\Lambda} (\Psi+Q) e^\Lambda\,,
\end{equation}
and we look for a gauge equivalent of the trivial solution $\Psi=0$.

It seems that this procedure cannot generate any non-trivial solutions.
However, this solution can become a physical one in several ways.
The first option is to have a finite radius of convergence with respect to
$\la$. By a rescaling of $\phi$ this value $\la_{\text{crit}}$ can be set
to unity. Then, for $|\la|<1$ the solution is indeed a
gauge solution, while for $|\la|>1$ it is not well
defined. For $\la=\pm 1$ the solution can either be
gauge solution, not be well defined, or be a physical solution.
Indeed, Schnabl's solution for the tachyon vacuum~\cite{Schnabl:2005gv}
is just of this form.
There, the solution at $\la=-1$ is not well defined, while the $\la=1$
case is the desired solution (after proper regularization).
A variant of this method would be to have $\la_{\text{crit}}=\infty$
such that the solution is well defined, but non-gauge at least in
one of the limits $\la\rightarrow \pm \infty$.
Another option would be to have some sort of a singular $\phi$,
such that the solution itself is regular. Then, $Q\phi$ is an exact
solution in a ``large Hilbert space'', but is
a nontrivial element of the cohomology when considering the smaller
Hilbert space.
This is the method that we want to employ.

The rest of the paper is organized as follows.
In section~\ref{sec:photon} we introduce
the analytical solution describing the photon marginal deformation
$\la c\partial X$. Then, in section~\ref{sec:gauge}, we discuss other
similar solutions and the issue of gauge equivalence. Our solution is
obtained using CFT methods. Therefore, we devote section~\ref{sec:coefficients}
to the oscillator form of the solution. Our construction is especially
useful for solutions whose OPE is singular, such as the photon marginal
deformation. However, it can be used also to describe other marginal
deformation. This issue is studied in section~\ref{OtherSolustion}.
Next, in section~\ref{sec:previous}
we comment on the relation of our solution to the previously found ones.
Finally, we present our conclusions in section~\ref{sec:conc}.

Following are some of the conventions we use in the paper.
Schnabl's solution is based on the wedge
states~\cite{Rastelli:2000iu,Furuuchi:2001df,Rastelli:2001vb,Schnabl:2002gg}
\begin{equation}
\psi_n = \ket{n+1} = \hat U_{n+1}\ket{0} = e^{\frac{1-n}{2}\hat\cL_0}\ket{0},
\end{equation}
where we use the notations
\begin{equation}
\hat U_n \equiv U^\dag_n U^\nodag_n\,,\qquad
\hat\cL_0 \equiv \cL^\nodag_0+\cL^\dag_0\,.
\end{equation}
Here, $\cL_0$ is the zero mode of the energy momentum tensor in the
coordinates
\begin{equation}
z=\tan^{-1}\xi\,,
\end{equation}
and $\cL_0^\dag$ is its conjugate.
The $\xi$ coordinate is the standard one, where the local coordinate patch
is half a unit circle.
The $z$ coordinate system is natural when working with insertions over the wedge
state $\ket{2}$, which is the $SL(2)$ invariant vacuum.
A natural generalization is to the coordinate system of the wedge state
$\ket{n}$
\begin{equation}
z^{(n)}= \frac{n}{2} z\,.
\end{equation}

Throughout the paper we work in the $z^{(n)}$ coordinates, when considering
states built as insertions over the wedge state $\ket{n}$. All operators, such
as $\partial X,c,...$ should be understood as defined in the
relevant coordinate system.
Also, throughout the paper we star-multiply string fields, keeping the star
product implicit. Similarly, a function of a string
field represents the Taylor expansion of the function, with products
given by star products.

\section{The photon marginal deformation}
\label{sec:photon}

Every marginal solution has a free parameter, which we mark as $\lambda$.
It is natural to expand the solution in orders of $\la$,
\begin{equation}
\Psi= \sum_{n=1}^\infty \lambda^n\psi_n\,.
\end{equation}
Plugging this into the equation of motion
and comparing terms with the same power of $\lambda$
gives the recursive relations
\begin{equation}
Q\psi_n = -\sum_{k=1}^{n-1}\psi_k\psi_{n-k}\,.
\end{equation}
For (formally) pure-gauge solutions~(\ref{eq:Psi}) we get
\begin{equation}
\label{eq:phi}
\psi_n = (Q\phi)\phi^{n-1}\,.
\end{equation}

We will attempt to generate the photon marginal solution
by making a singular choice of $\phi$
\begin{equation}
\label{initCond}
\phi = \epsilon_\mu X^\mu(0)\ket{0} = \epsilon_\mu x_0^\mu\ket{0}
    \quad \Rightarrow \quad
\psi_1 = -i \sqrt{2}\epsilon_\mu \alpha^\mu_{-1}c_1\ket{0}.
\end{equation}
Here, we work in the $\al'=1$ convention. Moreover, we
are interested in boundary operators.
Thus\footnote{
We shall henceforth omit the
polarization vector $\epsilon_\mu$ and the $\mu$ index over $X$
for simplicity.},
\begin{equation}
\partial X(z)_{\text{boundary}}=2\partial X(z)_{\text{bulk}}\,.
\end{equation}
In particular we have
\begin{equation}
\label{X0al}
\partial X(0)\ket{0}=-i\sqrt{2}\al_{-1}\ket{0}.
\end{equation}

The description of $\psi_1$ as a formal exact state~(\ref{initCond})
does not contradict the fact that $\psi_1$ is a physical state, i.e.
non-exact.
The reason is that $\phi$ is not in the Hilbert space,
since a state whose representation in position space is
\begin{equation}
\phi(x_0) = \braket{x_0}{\phi} = x_0\,,
\end{equation}
is not a normalizable state.
It is therefore clear that our choice of $\phi$ generates a non-trivial
solution.
Still, to show that this solution is legitimate we have to show that all
the $\psi_n$'s are regular, i.e. $x_0$ independent\footnote{An alternative
procedure, which we don't pursue in this work, would be the construction of
solutions with nontrivial but normalizable $x_0$ dependence.
We believe that this line of thought may be of help in the construction
of lump solutions.}.
The first order, $\psi_1$ is regular by construction.
The higher orders, however, have factors of $\phi^{n-1}$ in their
definition~(\ref{eq:Psi}) and are therefore singular,
with $\cO(x_0^{n-1})\ket{0}$ terms.
To correct these singularities we have to add counter terms to $\phi$,
i.e. terms of the form
\begin{equation}
\label{Xn}
\phi_n\approx X^n\,,
\end{equation}
where we need to specify the location of each $X$ insertion.
The gauge parameter $\phi$ becomes itself $\la$ dependent
\begin{equation}
\phi=\sum_{n=1}^\infty \la^{n-1}\phi_n\,.
\end{equation}
For expressions of the form~(\ref{Xn}) to make sense, they should be
normal ordered.
Henceforth, normal ordering of operator insertions at the
same point is implicit.
The relevant normal ordering scheme is boundary normal ordering.
In particular, all the variables that we use
are defined only on the boundary.

To calculate these counter terms
we need the commutation relation of $Q$ with $X(z)^n$
We use the relation
\begin{equation}
\label{Qexp}
[Q,e^{ipX(z)}]=\big(p^2\partial c e^{ipX}+ip c\partial X e^{ipX}\big)(z)\,,
\end{equation}
and derive it $n$ times to get
\begin{align}
\label{QX}
[Q,X^n] =& (-i\partial_p)^n[Q,e^{ipX}]\Big|_{p=0}
    = (-i)^n\left(2i^{n-2}\binom{n}{2}\partial c X^{n-2}
    + i(i)^{n-1}\binom{n}{1}c \partial X X^{n-1}\right)
\nonumber\\
    =& n c \partial X X^{n-1} - n(n-1)\partial c X^{n-2}\,.
\end{align}

Had we used~(\ref{eq:phi}) to construct the solution we would get
\begin{align}
\psi_2 &= (Q\phi_1) \phi_1=
    \hat U_3 c \partial X(-\frac{\pi}{4}) X(\frac{\pi}{4})
     \ket{0}.
\end{align}
This is obviously singular, in the sense that the coefficient of $x_0$ is
non-zero. To remedy this problem we add to $\phi$ a counter term of the form
\begin{align}
\phi_2=-\frac{1}{2}\hat U_3 X^2(-\frac{\pi}{4})\ket{0}.
\end{align}
This gives a regular solution
\begin{align}
\nonumber
\label{psi2}
\psi_2 &= (Q\phi_1) \phi_1+Q\phi_2=
    \hat U_3 \Big(c \partial X(-\frac{\pi}{4})
       \big(X(\frac{\pi}{4})-X(-\frac{\pi}{4})\big)+
         \partial c(-\frac{\pi}{4})\Big)
     \ket{0}\\
 &\equiv (c\partial X,X)-(c\partial X X,1)+(\partial c,1)=
   (c\partial X,1)\big((1,X)-(X,1)\big)+(\partial c,1)\,.
\end{align}
Here, a new notation has been introduced. As we consider here
the wedge state $\ket{3}$ (formed by acting with $\hat U_3$ on the vacuum)
with two possible insertion sites, we specify the
insertions in these sites in a vector form, where the $1$ stands for the
identity insertion, i.e., no insertion.
The multiplication in the last equality acts point wise.

At order $n$ we shall
have to deal with $n$ insertions over the wedge state $\ket{n+1}$,
symmetrically distributed around $z^{(n+1)}=0$, with a distance of
$\frac{\pi}{2}$ between two consecutive insertion sites.
There, we use an $n$-vector notation. Since the wedge state number is
correlated with the number of insertion sites this notation is unambiguous.

The expression for $\psi_2$ seems regular since there is no $x_0$
dependence in $(1,X)-(X,1)$, but this analysis is a bit too naive.
One has to remember that we have normal ordering only for operators
at the same site. Therefore,
\begin{equation}
(c\partial X,1)\big((1,X)-(X,1)\big) =
    :c\partial X(-\frac{\pi}{4})::X(\frac{\pi}{4}):\ket{0}-
    :c\partial X X(-\frac{\pi}{4}):\ket{0}.
\end{equation}
It is easy to normal order this term and see that there is no $x_0$
dependence.
This case is trivial since the new term that appears in the normal ordering
has no $X$ dependence.
For higher order terms, normal ordering $X^n$ will introduce
$X^{n-2}$ and lower order operators.
However, let $P(X_1,..,X_n)$ be a polynomial in $X_i$,
such that in each monomial operators at different sites are ordered
from left to right.
If $P$ is $x_0$ independent,
\begin{align}
\partial_{x_0}P=\sum_{k=1}^n \partial_{X_k}P=0\,,
\end{align}
then the normal ordered expression given by
\begin{align}
\label{NormalOrder}
:P: = e^{\frac{1}{2}\sum_{i\ne j}f_{i,j}^{(n)}\partial_{X_i}\partial_{X_j}+
 \sum_{i\neq j}g_{i,j}^{(n)}\partial_{(\partial X_i)}\partial_{X_j}}P\,,
\end{align}
is also $x_0$ independent.
This stems from the fact that the normal ordering operator inside
the exponent commutes with the operator $\sum_{k=1}^n \partial_{X_k}$.
This result is independent of the
form of the structure functions $f_{i,j}^{(n)},g_{i,j}^{(n)}$, as long as
they are $X_i$-independent\footnote{
While one can ignore the normal ordering issues for proving
$x_0$-independence, they are relevant in some computations.
The explicit structure functions will be needed for writing down the
fully normal ordered solution.}.
Notice that the fully normal ordered polynomial $:P:$ can be manipulated
as a standard polynomial, since all of its components commute with each other.

We now continue with the third order,
\begin{align}
\nonumber
\psi_3 &= (Q\phi_1)\phi_1 \phi_1+(Q\phi_2)\phi_1+(Q\phi_1)\phi_2+Q\phi_3\\
 &=(c\partial X,1,1)\big((1,X,X)-(X,1,X)-\frac{1}{2}(1,X^2,1)\big)
    +(\partial c,1,X)+Q\phi_3\,.
\end{align}
The counter term $\phi_3$ is obviously needed. There is an $X(0)$ insertion
multiplying $\partial c(-\frac{\pi}{2})$, and the coefficient of the
$(c\partial X)(-\frac{\pi}{2})$ insertion 
also has an $x_0^2$ dependence.
However, by setting
\begin{align}
\phi_3=\frac{1}{6}\hat U_4 X^3(-\frac{\pi}{2})\ket{0}=
  \frac{1}{6}(X^3,1,1)\,,
\end{align}
we are led to
\begin{align}
\nonumber
\psi_3 &= Q\phi_1 \phi_1 \phi_1+Q\phi_2 \phi_1+Q\phi_1 \phi_2+Q\phi_3\\
 &=(c\partial X,1,1)\big((1,X,X)-(X,1,X)-\frac{1}{2}(1,X^2,1)
   +\frac{1}{2}(X^2,1,1)\big)\\
\nonumber
    &+(\partial c,1,1)\big((1,1,X)-(X,1,1)\big)\,.
\end{align}
The coefficient of $\partial c$ is obviously regular and it is
also clear that there is no $x_0^2$ term in the coefficient of $c\partial X$.
There could have been a linear $x_0$ term there, though.
In order to see that it is absent, we write $X$ at site $k$, as
$x_0+\tilde X_k$, where $\tilde X_k$ is the regular part of the operator.
Then, we see that
\begin{align}
\nonumber
&(x_0+\tilde X_2) (x_0+\tilde X_3)-(x_0+\tilde X_3) (x_0+\tilde X_1)+
\frac{1}{2} (x_0+\tilde X_1)^2-\frac{1}{2}(x_0+\tilde X_2)^2\\
&=\tilde X_2 \tilde X_3-\tilde X_3 \tilde X_1+
  \frac{\tilde X_1^2}{2}-\frac{\tilde X_2^2}{2}\,,
\end{align}
so the linear term drops out as well, and the result is regular.

Continuing to higher orders, we guess that the general form of the
counter terms is
\begin{align}
\label{phiForm}
\phi_n=\frac{1}{n!}(-1)^{n-1}(X^n,1,\ldots,1)\,.
\end{align}
With this ansatz we have to check as before that all the
$x_0$ coefficients vanish, without considering the issue of normal ordering.
We prove this regularity condition by induction.
Suppose that $\psi_k$ is regular for all $k<n$.
The state $\psi_n$ is composed of all ways to partition the $n$ sites among
the various $\phi_k$'s, with $Q$ acting on the first site.
We concentrate on the last $\phi$ in any given partition
and write $\psi_n$ in terms of the $\psi_k$'s as
\begin{align}
\label{decomposePsi}
\psi_n=Q\phi_n+\sum_{k=1}^{n-1} (\psi_{n-k},\phi_k)\,.
\end{align}
We introduced a new notation here of composing two vectors into one
longer vector. This is exactly the operation of the star product.
Now,
\begin{align}
\partial_{x_0}\psi_n=Q\partial_{x_0} \phi_n+
   \sum_{k=1}^{n-1} \Big((\partial_{x_0}\psi_{n-k},\phi_k)
   +(\psi_{n-k},\partial_{x_0}\phi_k)\Big)\,.
\end{align}
The first term inside the sum drops out according to the induction hypothesis,
while from~(\ref{phiForm}) we see that
\begin{align}
\label{recRelation}
\partial_{x_0}\phi_k=-(\phi_{k-1},1)\,.
\end{align}
Thus, we are left with
\begin{align}
\nonumber
\partial_{x_0}\psi_n=&-\Big(Q (\phi_{n-1},1)-(\psi_{n-1},1)
   +\sum_{k=2}^{n-1} (\psi_{n-k},\phi_{k-1},1)\Big)\\
 =&-\big((\psi_{n-1},1)-(\psi_{n-1},1)\big)=0\,,
\end{align}
where in the first equality we separated the $k=1$ term and in the second equality
composed the first term with the sum as in~(\ref{decomposePsi}).
This completes the proof.

\section{Gauge-choice independence}
\label{sec:gauge}

In this section we generalize our construction and find a large family of
solutions. All these solutions should be gauge equivalent.
Thus, we study the gauge equivalence of these solutions.
First, in subsection~\ref{GaugeNoCT} we consider gauge freedom ignoring
counter terms.
There, $\phi$ represents a given formal gauge generator
of a specific marginal deformation, not necessarily the photon. As already
mentioned, this construction fails to produce well defined solutions without
introducing counter terms.
In subsection~\ref{GaugeWithCT} we study the consequences of adding them
for the photon marginal deformation. 

\subsection{Ignoring counter terms}
\label{GaugeNoCT}

It is clear that the linearized equation of motion $\psi_1=Q\phi$
does not uniquely
define the solution to be of the form~(\ref{eq:Psi}),
in which the $Q$ operates only on the leftmost $\phi$ at each
order~(\ref{eq:phi}).
For example, we could have considered instead
\begin{equation}
\label{PsiR}
\Psi_R = \frac{\la}{1+\la \phi} Q\phi\,,
\end{equation}
which is generated by the gauge field
\begin{equation}
\Lambda = \log(1+\lambda\phi)\,.
\end{equation}
Here, $Q$ acts only on the rightmost $\phi$.

Do these solutions describe the same physical state?
When several marginal directions exist, the non-linear (that is, finite)
extension of the initial infinitesimal deformation is not unique, and
can ``point'' in various (possibly $\la$-dependent) directions in the
vector space of infinitesimal marginal deformations.
However, it seems that in our construction the only marginal deformation
considered is related to the zero mode of a single $X$ coordinate. Thus,
all the solutions should be gauge equivalent.
Indeed, at the second order, the difference between the two solutions is
\begin{equation}
\label{psiLpsiR}
\psi_2^L-\psi_2^R=Q(\phi^2)\,.
\end{equation}
So they are indeed gauge equivalent to this order.

One can also consider other extensions of the solution.
If we restrict ourselves to solutions, which depend only on $\phi$ and
have a total $n^{th}$ power of $\phi$ at order $n$, then
a general ansatz can be written as
\begin{equation}
\label{oneGaugePar}
\Psi=\sum_{n=1}^\infty \la^n \sum_{k=1}^{n} \gamma_{n,k} \phi^{k-1} (Q\phi) \phi^{n-k}\,,
\end{equation}
and the requirement that $\Psi$ is a solution imposes restrictions on the
values of the coefficients $\gamma_{n,k}$. The general solution for
$\gamma_{n,k}$ amounts to the freedom of adding $Q(\phi^n)$ at order $n$
to the solution. This is a straightforward generalization
of~(\ref{psiLpsiR}), with one gauge parameter at each order.

Another way to represent the gauge freedom is by specifying $\La$.
Any $\La$ that agrees to first order with the
original one and depends only on powers of $\phi$ should be equivalent to it.
Such a $\La$ has a Taylor expansion
\begin{equation}
\La=\la \phi+\sum_{n=2}^\infty a_n(\la \phi)^n\,,\qquad
e^\La=1+\la \phi+\sum_{n=2}^\infty b_n(\la \phi)^n\,,
\end{equation}
with obvious relations between the coefficients $a_n,b_n$.
Two solutions which agree up to order $n-1$, differ by
\begin{equation}
\label{LaExpansion}
e^{-\La_1}Q e^{\La_1}-e^{-\La_2}Q e^{\La_2}=
  \la^n (a^{(1)}_n-a^{(2)}_n)Q(\phi^n)+\cO(\la^{n+1})=
  \la^n (b^{(1)}_n-b^{(2)}_n)Q(\phi^n)+\cO(\la^{n+1})\,.
\end{equation}
Thus, we see that choosing the expansion coefficients $a_n$ or $b_n$
is equivalent to choosing $\gamma_{n,k}$ consistently.
In particular it is easy to see that
\begin{equation}
b_n=\gamma_{n,1}\,.
\end{equation}

The two simple solutions described above correspond to
\begin{equation}
\gamma_{n,k}^L=\delta_{k,1}\,,\qquad \gamma_{n,k}^R=(-1)^{n-1}\delta_{k,n}\,.
\end{equation}
The gauge freedom described above allows for a more symmetric solutions,
such as
\begin{equation}
\gamma_{n,k}^S=\frac{1}{2^{n-1}}(-1)^{k-1}\,,
\end{equation}
which is generated by
\begin{equation}
\La=\log\Big(\frac{1+\frac{\la \phi}{2}}{1-\frac{\la \phi}{2}}\Big)
=2\tanh^{-1}\Big(\frac{\la \phi}{2}\Big)
\,.
\end{equation}
Another, relatively symmetric combination that we found is
\begin{equation}
\gamma_{n,k}=\left\{
\begin{array}{ccccccccccccccccccc}&&&&&&&&&1\\
&&&&&&&&\frac{1}{2} && -\frac{1}{2}\\
&&&&&&&\frac{1}{2} && 0 && \frac{1}{2}\\
&&&&&&\frac{3}{8} && -\frac{1}{8} && \frac{1}{8} && -\frac{3}{8}\\
&&&&&\frac{3}{8} && 0 && \frac{1}{4} && 0 && \frac{3}{8}\\
&&&&\frac{5}{16} && -\frac{1}{16} && \frac{1}{8} && -\frac{1}{8} && \frac{1}{16}
   && -\frac{5}{16} \\
&&&\frac{5}{16} && 0 && \frac{3}{16} && 0 && \frac{3}{16} && 0 && \frac{5}{16}\\
&&\frac{35}{128} && -\frac{5}{128} && \frac{15}{128} && -\frac{9}{128}
   && \frac{9}{128} && -\frac{15}{128} && \frac{5}{128} && -\frac{35}{128} \\
&\frac{35}{128} && 0 && \frac{5}{32} && 0 && \frac{9}{64} && 0
   && \frac{5}{32} && 0 && \frac{35}{128}\\
\iddots &&&&&&&&& \cdots &&&&&&&&& \ddots
\end{array}
\right.
\end{equation}
This expansion is generated by
\begin{equation}
\La=\frac12\log\left(\frac{1+\la \phi}{1-\la \phi}\right)
=\tanh^{-1}(\la \phi)\,,
\end{equation}
and the expansion coefficients have the peculiar property
\begin{equation}
\sum_{k=1}^n \gamma_{n,k}=\left\{
    \begin{array}{cc}0 & \qquad n\equiv_2 0\\
                     1 & \qquad n\equiv_2 1 \end{array}\right.\,.
\end{equation}

With the richness of gauge descriptions one may wonder whether they all
have the same range of validity in parameter space. Here, we have a single
parameter $\la$, and it is clear on physical grounds that its value can be
arbitrary. However, since we should not expect a gauge choice to be globally
valid, it is possible that as we increase $\la$ the boundary of validity
is attained for some of the gauge choices.
There may be a subtlety here, since there can be a non-trivial relation
\begin{equation}
\label{CFTSFT}
\la_{SFT}=f(\la_{CFT})=\la_{CFT}+\ldots\,,
\end{equation}
with $f$ a monotonic function.
It can happen that as $\la_{CFT}$ goes to infinity, $\la_{SFT}$ reaches
a finite value. If this is the case, it can happen that the
radius of convergence with respect to $\la_{SFT}$ would always be finite.
Also, note that the function $f$ generally depends on the solution.
We saw that for our ansatz this reparametrization is absent for the
leading order~(\ref{LaExpansion}), but it can appear at higher ones.
We assume for now that this does not happen.

As all the gauge choices that we explicitly considered so far are described by
functions of $\la\phi$ with a finite radius of convergence, it seems plausible
that the gauge choice would break down at some stage\footnote{One should keep
in mind, though, that the relevant product is the star product. So the series
expansion should break down when an eigenvalue in some relevant space with
respect to this product attends a critical value.}.
One may hope that a wider range of validity is attained
by choosing a function $\La$ such that $e^{\pm \La}$ are complete.
The simplest example of the form~(\ref{LaExpansion}) is
\begin{equation}
\label{expSol}
\La=\la \phi\,.
\end{equation}
With this choice one gets
\begin{equation}
\gamma_{n,k}=\frac{(-1)^{k-1}}{n (k-1)! (n-k)!}\,.
\end{equation}
These coefficients have the peculiar property that
\begin{equation}
\forall n>1\qquad \sum_{k=1}^n \gamma_{n,k}=0\,.
\end{equation}

\subsection{Including counter terms}
\label{GaugeWithCT}

In the previous subsection we have been ignoring counter terms.
Now, we want to address the question of the uniqueness of the solution
including the counter terms for the case of the photon marginal deformation.
Any two solutions that are the same up to a given order
$\Psi_1-\Psi_2=\cO(\la^{n})$,
are gauge equivalent at the next order, provided that their difference
is $Q$-exact to this order,
$\Psi_1-\Psi_2=\la^{n}Q(\Upsilon)+\cO(\la^{n+1})$,
as in~(\ref{LaExpansion}).
Recall that $Q$ acting on a singular expression should not be considered
as $Q$-exact,
but only as $Q$-closed.
In this subsection we study the gauge freedom at the leading order
and find the full (all order) gauge transformation between $\Psi_L$
and $\Psi_R$.

With the counter terms taken into account, the two simple solutions 
$\Psi_L,\Psi_R$ defined by equations (\ref{eq:Psi}) and (\ref{PsiR})
are generated by
\begin{align}
\phi_L &= (X)-\frac{\la}{2!} (X^2,1)+\frac{\la^2}{3!} (X^3,1,1)+...+
\frac{(-\la)^{n-1}}{n!} (X^n,1,...,1)+...\,,\\
\phi_R &= (X)+\frac{\la}{2!} (1,X^2)+\frac{\la^2}{3!} (1,1,X^3)+...+
\frac{\la^{n-1}}{n!} (1,...,1,X^n)+...\,.
\end{align}
The difference between the solutions $\Psi_L-\Psi_R$, before
counter terms are taken into account is given at the second order
by~(\ref{psiLpsiR})
\begin{equation}
\Delta_2=Q(\phi^2)=Q(X,X)\,.
\end{equation}
This is obviously $Q$ acting on a singular expression.
However, with counter terms taken into account the difference is in fact
\begin{equation}
\Delta_2=Q\big((X,X)-\frac{1}{2}(X^2,1)-\frac{1}{2}(1,X^2)\big)\,,
\end{equation}
which is $Q$ on a regular expression.
So the solutions are gauge equivalent at this order.
This is the most general regular expression at this order, which is
of the form of our ansatz.

The most general third order difference, between solutions of the form we
consider here, that agree up to the second order, is
\begin{align}
\Delta_3=& Q \phi_3\\
\nonumber
\phi_3=& \al_{1,1,1}(X,X,X)+
  \al_{3,0,0}(X^3,1,1)+\al_{0,3,0}(1,X^3,1)+\al_{0,0,3}(1,1,X^3)\\
+&  \al_{2,1,0}(X^2,X,1)+\al_{2,0,1}(X^2,1,X)+
    \al_{1,2,0}(X,X^2,1)+\al_{0,2,1}(1,X^2,X)\\
\nonumber
+&  \al_{1,0,2}(X,1,X^2)+\al_{0,1,2}(1,X,X^2)\,.
\end{align}
The ten parameters $\al_{i,j,k}$ should be chosen such that $\Delta_3$
is regular.
To that end, we should require that the coefficients of
$c\partial X x_0^2$, $c\partial X x_0 \tilde X_i$
and $\partial c x_0$ in the three sites should be zero.
Thus, there are totally $3+3\times 3+3=15$ linear equations restricting
the values of the $\al_{i,j,k}$'s. Twelve homogeneous linear equations
in ten variables are of course dependent.
It turns out that the general solution forms a four-dimensional space.
That we have here four free parameters, whereas only a single
gauge parameter was advocated in~(\ref{oneGaugePar}), stems from the fact
that we are dealing here with a more general ansatz
due to the appearance of counter terms.
Now, to verify that all the solutions are the same also at the third order,
we have to check the form of $\phi_3$ in this four-dimensional space.
It turns out that it is indeed regular.

In a similar way at order $n$ we consider
\begin{align}
\Delta_n=& Q\phi_n\,,\qquad\qquad
\phi_n=\sum_{\stackrel{i_1,..,i_n=0}{_{i_1+..+i_n=n}}}^n
   \al_{i_1,..,i_n}(X^{i_1},..,X^{i_n})\,.
\end{align}
Here, the number of coefficients is
\begin{align}
\#_{\al_{\vec{i}}}=\binom{2n-1}{n}\,.
\end{align}
Let us note that
there are restrictions coming from the expansion of
$c\partial X P_{n-1}(\tilde X_i, x_0)$ and $\partial c P_{n-2}(\tilde X_i, x_0)$,
in the $n$ sites, where $P_{n-1}$ and $P_{n-2}$ are homogeneous polynomials of
degrees $n-1$, $n-2$ respectively. A rank-$k$ homogeneous polynomial with $m$
variables is composed of
\begin{equation}
C_{k,m}\equiv\binom{k+m-1}{k}
\label{Ckm}
\end{equation}
monomials.
The constraints we have amount
to requiring that the two polynomials depend only on $n$ out of their
$n+1$ variables.
Thus, the total number of equations is
\begin{align}
\#_{\text{eq}}=n
\Big((C_{n-1,n+1}-C_{n-1,n})+(C_{n-2,n+1}-C_{n-2,n})\Big)
   =(3n-4)\binom{2n-3}{n-1}\,.
\end{align}
For $n>3$ the number of equations keeps being larger than the number of variables,
just as is the case for $n=3$.
At the fourth order only $20$ out of a total of $80$ equations
(in $35$ variables) are independent.
Thus, the space of allowed $\phi_4$ is
$35-20=15$-dimensional.
It turns out that when restricted to this space of solutions
$\phi_4$ is regular.

We want to prove this result for the general case, that is, we want to show,
that given $\al_{\vec i}$ such that $Q\phi_n$ is regular, so is
$\phi_n$ itself. In the expression
\begin{align}
Q\phi_n=\sum_{\stackrel{i_1,..,i_n=0}{_{i_1+..+i_n=n}}}^n
   \al_{i_1,..,i_n}\sum_{k=1}^n
 (X^{i_1},..,i_k c\partial X X^{i_k-1}-i_k(i_k-1)\partial c X^{i_k-2},..,X^{i_n})\,,
\end{align}
we have to demand that the coefficient of each $c\partial X$ and each
$\partial c$, is $x_0$ independent. We write the coefficient of $c\partial X$
at site $k$ as
\begin{align}
\xi_k=\partial_{X_k}\sum_{\stackrel{i_1,..,i_n=0}{_{i_1+..+i_n=n}}}^n
   \al_{i_1,..,i_n}
 (X^{i_1},..,X^{i_n})=\partial_{X_k} \phi_n\,,
\end{align}
while that of $\partial c$ can be written as
\begin{align}
\zeta_k=-\partial_{X_k}^2\sum_{\stackrel{i_1,..,i_n=0}{_{i_1+..+i_n=n}}}^n
   \al_{i_1,..,i_n}
 (X^{i_1},..,X^{i_n})=-\partial_{X_k}^2 \phi_n\,,
\end{align}
It is clear that $\partial_{x_0} \xi_k=0$ implies $\partial_{x_0} \zeta_k=0$,
so it is enough to consider the former and what we have to show is that
\begin{align}
\partial_{x_0}\xi_k=\sum_{i=1}^n\partial_{X_i}\xi_k=0\quad \forall k
   \quad\Rightarrow \quad\partial_{x_0}\phi_n=\sum_{k=1}^n \xi_k=0\,.
\end{align}
Now, consider
\begin{align}
\nonumber
0=&\sum_{k=1}^n X_k \sum_{i=1}^n\partial_{X_i}\xi_k=
 \sum_{i=1}^n\partial_{X_i}\sum_{k=1}^n X_k \partial_{X_k}\phi_n-
   \sum_{k=1}^n \partial_{X_k}\phi_n\\
=& (n-1)\sum_{i=1}^n \partial_{X_i}\phi_n=(n-1)\partial_{x_0}\phi_n\,,
\end{align}
where we used Euler's homogeneous function theorem.
We see that for $n>1$ regularity of the coefficients indeed implies that of
$\phi_n$, so all the solutions are indeed gauge equivalent,
at least up to the first order where they differ.

It is clear that any gauge field $\phi$ that is $x_0$ independent produces
an $x_0$ independent $Q\phi$. We have just prove also the opposite direction.
Thus, we can now characterize the $\phi$ space in this way. This enables us
to calculate the dimension of this space, since for a homogeneous polynomial
of degree $n$,
\begin{equation}
\partial_{x_0} P(x_k)=\sum \partial_{X_i} P(x_k)=0 \quad \Rightarrow \quad
P=\sum k_{i_2,\ldots,i_n}(x_1-x_2)^{i_2}\cdots(x_1-x_n)^{i_n}\,,
\end{equation}
with $k_{i_2,\ldots,i_n}$ a set of coefficients. From~(\ref{Ckm}), we see
that
\begin{equation}
\dim(\phi_n)=\binom{2n-2}{n}\,,
\end{equation}
which agrees with the quoted results for $n<5$. Moreover, we see that
the most general solution can be found from~(\ref{recRelation}), while in
this subsection we studied the homogeneous version of this equation.

One of the gauge degrees of freedom is related to a change in $\gamma_{n,k}$.
This is generated by $Q(\phi^n)$, that is, by $(X,\ldots,X)$. When one wants
to consider a change in the gauge within a choice of $\gamma_{n,k}$ (or $\La$),
one should constrain the gauge parameter to the orthogonal space to this
direction, say, by setting the coefficient of $(X,\ldots,X)$ to zero. However,
it is not clear to us in which sense the inner product that we use in the
space of $\al_{\vec{i}}$'s is canonical. Thus, this choice of
``orthogonal direction'' is somewhat arbitrary.

This brings us back to the discussion regarding the range of validity of
the different gauge choices. It is hard to disentangle the choice of counter
terms from the choice of $\gamma_{n,k}$. Moreover, with counter terms taken
into account, the gauge parameter $\phi$ is itself $\la$ dependent. Thus,
it is possible that the solutions $\Psi_{L,R}$ have an infinite radius of
convergence. It can also happen that the solution~(\ref{expSol}) would have
a finite radius of convergence with an appropriately chosen set of counter terms
and the radius can also depend on the specific choice.
The easiest way in practice to generate an appropriate set of counter terms
for a given $\gamma_{n,k}$, is by imposing~(\ref{recRelation}),
while fixing the coefficients $\al_{1,\ldots,1}$ that generate the specific
choice of $\gamma_{n,k}$.
We try to estimate
the radius of convergence of the solution $\Psi_L$ in
section~\ref{sec:coefficients}.

We now want to prove that the two solutions that we found explicitly
$\Psi_L,\Psi_R$, which differ at the second order,
are indeed exactly gauge equivalent.
To that end we first note that~(\ref{generalGauge}) implies that when two
finite gauge transformations are combined,
the resulting transformation is given by
\begin{align}
e^\La=e^{\La_{1}}e^{\La_{2}}\,.
\end{align}
We now combine the singular gauge transformations that send $\Psi_L$ to the
trivial solution and the trivial solution to $\Psi_R$, to get the gauge
transformation from $\Psi_L$ to $\Psi_R$,
\begin{align}
\nonumber
e^\La=& e^{-\La_L}e^{\La_R}=(1-\la\phi_L)(1+\la\phi_R)=
  \sum_{n=0}^\infty \frac{(-\la)^n}{n!}\sum_{m=0}^\infty \frac{\la^m}{m!}
     \underbrace{(X^n,1,\ldots,1,X^m)}_{n+m}\\
=&\sum_{n=0}^\infty \la^n \sum_{m=0}^n \frac{(-1)^m}{(n-m)!m!}
     X_1^m X_n^{n-m}=\sum_{n=0}^\infty \frac{\la^n}{n!}(X_n-X_1)^n\,,
\end{align}
where in the second line we suppressed the irrelevant sites. However, one
should remember that the expression at order $n$ is defined over the wedge
state $\ket{n}$.
We see that the resulting transformation is manifestly regular, so $\Psi_{L,R}$
are truly gauge equivalent to all orders.

\section{Evaluating coefficients in the oscillator representation}
\label{sec:coefficients}

Next we want to calculate the coefficients of the different
fields in the oscillatory expansion for the photon marginal solution.
This requires normal ordering our expression.

The easiest way to work with the (non-primary) $X^n$ fields is by their
defining expression
\begin{equation}
X^n(z)\equiv (-i\partial_p)^n e^{ipX(z)}\Big|_{p=0}\,.
\end{equation}
These fields are normal ordered in the sense of the $z$ coordinates.
To find the OPE of two $X(z^{(n)})$ operators we recall first that in the $\xi$
plane
\begin{equation}
\partial X(\xi_1)\partial X(\xi_2)=:\partial X(\xi_1)\partial X(\xi_2):
   -\frac{2}{\xi_{12}^2}\,,
\end{equation}
where the factor of 2 comes from the fact that we are working with
boundary operators as in~(\ref{X0al}).
The conformal transformation
\begin{equation}
z^{(n)}=\frac{n}{2}\tan^{-1}\xi
\end{equation}
gives
\begin{equation}
\label{dXdXNO}
\partial X(z^{(n)}_1)\partial X(z^{(n)}_2)=
  :\partial X(z^{(n)}_1)\partial X(z^{(n)}_2):
   -\Big(\frac{2}{n}\Big)^2\frac{2}{\sin^2(\frac{2}{n}z_{12}^{(n)})}\,.
\end{equation}
Integrating we get
\begin{align}
\label{dX_X_zn}
\partial X(z_1^{(n)}) X(z_2^{(n)})=&:\partial X(z_1^{(n)}) X(z_2^{(n)}):
   -\frac{4}{n}\cot(\frac{2}{n} z_{12}^{(n)})\,,\\
\label{X_X_zn}
X(z_1^{(n)}) X(z_2^{(n)})=&:X(z_1^{(n)}) X(z_2^{(n)}):
   -2\log\Big(\sin\Big|\frac{2}{n}z_{12}^{(n)}\Big|\Big)\,
\end{align}
Other than giving the form of the singular part of the OPE, these equations
also define what we mean by normal ordering in the $z^{(n)}$-plane.

The $X^m$ operators are not primaries. We would need their transformation
properties for going between the $z^{(n)}$ and the $\xi$ planes.
From their definition, it follows that
\begin{align}
\nonumber
X^m(\xi)&=(-i\partial_p)^m e^{i p X(\xi)}\Big|_{p=0}=
  (-i\partial_p)^m \bigg(\Big(\frac{n}{2}\cos^2(z^{(n)})\Big)^{p^2}
      e^{i p X(z^{(n)})}\bigg)\bigg|_{p=0}\\
&= \sum_{j=0}^{\lfloor \frac{m}{2} \rfloor}\frac{m!}{(m-2j)!j!}
   X^{m-2j}(z^{(n)})\log^j\Big(\frac{2}{n\cos^2(z^{(n)})}\Big)\nonumber\\
&= e^{-\log\big(\frac{n}{2}\cos^2(z^{(n)})\big)\partial_X^2}
   \,X^m(z^{(n)})\,,
\label{X_xi_z}
\end{align}
where in the second equality we used Leibnitz' formula, as well as the
fact that only an even number of $p$-derivatives on $c^{p^2}$ gives
a non-zero result at $p=0$. Hence, also the floor-function on the sum.
Similarly,
\begin{align}
X^m(z^{(n)})
&= \sum_{j=0}^{\lfloor \frac{m}{2} \rfloor}\frac{m!}{(m-2j)!j!}
   X^{m-2j}(\xi)\log^j\Big(\frac{n}{2(1+\xi^2)}\Big)\nonumber\\
&= e^{-\log\big(\frac{2}{n}(1+\xi^2)\big)\partial_X^2}\, X^m(\xi)\,.
\label{X_z_xi}
\end{align}
Equations (\ref{X_xi_z}) and (\ref{X_z_xi}) relate powers of $X$ at the
same point that are normal ordered in the sense of the $\xi$
coordinates to those that are normal ordered in the sense of
the $z^{(n)}$ coordinates, and vice-versa (we note again that normal
ordering is taken to be implicit for operators inserted at the same
point). We can now use an argument analogous to the one
following~(\ref{NormalOrder}) to show that our construction, which leads to
$x_0$-independence in the sense of the $z^{(n)}$ coordinates,
automatically implies $x_0$-independence in the sense of the $\xi$
coordinates. Indeed, noting that
\begin{equation}
\partial_{x_0}X^m=\partial_X X^m\,,
\end{equation}
the fact that the final expression of (\ref{X_z_xi}) involves the
exponent of an operator independent of $m$ commuting with $\partial_X$
shows that $x_0$-independent expressions constructed using the
$z^{(n)}$ coordinates translate to $x_0$-independent expressions in
the sense of the $\xi$ coordinates.

We also need the normal ordering coefficients for~(\ref{NormalOrder}).
In the $z^{(n)}$ coordinate they are
\begin{align}
f_{i,j}^{(n)}=-2\log\Big(\sin\Big|\frac{2}{n}z_{ij}^{(n)}\Big|\Big)\,,\qquad
g_{i,j}^{(n)}=-\frac{4}{n}\cot(\frac{2}{n} z_{ij}^{(n)})\,,
\end{align}
while in the upper half plane
\begin{align}
f_{i,j}^{\text{UHP}}=-2\log|\xi_{ij}|
\,,\qquad
g_{i,j}^{\text{UHP}}=-\frac{2}{\xi_{ij}}\,.
\end{align}

We want to calculate the coefficients of the lowest state, i.e. the
tachyon state $c_1\ket{0}$, as well as that of the photon in the direction
in which we induce the marginal deformation, i.e.
the state $\al_{-1}c_1\ket{0}$.
This corresponds to calculating the expectation values
\begin{align}
\label{TachyonDef}
T(\la) &= \sum_{n=1}^\infty \la^n T_n\,,\qquad \,
  T_n= \frac{2}{n+1}\left<\partial c c, \psi_n\right>_{n+1},\\
A(\la) &= \sum_{n=1}^\infty \la^n A_n\,,\qquad
  A_n= \frac{i}{\sqrt{2}}\left<\partial c c \partial X,
     \psi_n\right>_{n+1},
\end{align}
where the prefactor $\frac{n+1}{2}$ in the first equation
comes from the conformal transformation
to $C_{\frac{(n+1)\pi}{2}}$, the cylinder with circumference
$\frac{(n+1)\pi}{2}$ and in the second expression we took into
account~(\ref{X0al}).
Symmetry dictates that for $T(\la)$ only the even powers are nontrivial, while
for $A(\la)$ only the odd powers are non trivial.
 
For the ghost sector the only correlators that we need are
\begin{align}
\big<\partial c c,c, \underbrace{1,\ldots,1}_{n-2}\big>
    &= \frac{n^2}{4}\sin^2(\frac{\pi}{n})\,,\\
\big<\partial c c,\partial c, \underbrace{1,\ldots,1}_{n-2}\big>
    &= \frac{n}{2}\sin(\frac{2\pi}{n})\,.
\end{align}
Here we introduced a new notation for the expectation value of a wedge
state $n$ with the $n$ possible canonical insertion points.
For the matter sector only the constants
that arise from the normal ordering procedure
as well as expressions that are a priori
$X$ independent\footnote{The only such case is the operator $\partial c$
that arises at the second order.} can contribute.

The first coefficient is\footnote{We work in conventions in which $\lambda$
is imaginary. Thus, odd coefficients appear to be imaginary.}
\begin{equation}
\label{A1}
A_1=\frac{i}{\sqrt{2}}\left<\partial c c,c \right>
    \left<\partial X,\partial X \right>
    =-i\sqrt{2}\,,
\end{equation}
where we used~(\ref{dXdXNO}). This result is trivial, since this is nothing
but the initial condition for the solution~(\ref{initCond}).
Next, we have
\begin{equation}
T_2=\frac{2}{3}\Big(\left<\partial c c,c,1 \right>
\big(\left<1,\partial X,X \right> - \left<1,\partial X X,1 \right> \big)
    +\left<\partial c c,\partial c,1 \right> \Big)\,,
\end{equation}
where we substituted the result for $\psi_2$~(\ref{psi2}).
The second term does not contribute, since it contains only $X$ insertions
at the same point, which are already normal ordered. The first term contributes
the normal ordering constant and the third term is the unique case, where
there is no $X$ dependence. All in all we get
\begin{equation}
T_2=\frac{2}{3}\Big(\frac{9}{4}\sin^2(\frac{\pi}{3})
(-\frac{4}{3})\cot\big(\frac{2}{3}(-\frac{\pi}{2})\big)
    +\frac{3}{2}\sin(\frac{2\pi}{3})\Big)=\sqrt{3}\,.
\end{equation}
Here, we used~(\ref{dX_X_zn}).

In a similar manner we can calculate higher order terms.
The only limitation is that the number of terms increases very fast
and the analytic expressions become extremely cumbersome.
We calculated the following values numerically:
\begin{align}
\nonumber
&T_4=1.39848,& &T_6=0.863328,& &T_8=0.477264,\\
&T_{10}=0.243907,& &T_{12}=0.117105,& &T_{14}=0.0534555,\\
\nonumber
&A_3=-5.2229i,& &A_5=-6.61142i,& &A_7=-5.84034i,\\
&A_9=-4.23182i,& &A_{11}=-2.67936i,& &A_{13}=-1.53481i.
\end{align}
These numbers seem to indicate that we have a finite radius of convergence.

A different choice of counter terms would have produced different
coefficients. It seems that the number of degrees of freedom in choosing the
counter terms grows fast with the level. Thus, one may hope that it would
be possible to choose them in a way that would minimize the coefficient
at each level. One has to remember though, that there are also an
infinity of other coefficients, that we did not evaluate.
For the series to converge, we have to impose convergence on the whole
infinite set, and a gauge that minimizes one coefficient can be less adequate
for minimizing the others. This issue deserves further study.

Note, however, that the change of a gauge at level $n$ cannot modify the
coefficient of $A_n$.
This results from the closeness of the bra state used for extracting the
photon coefficient. Let the gauge change be given in the leading order
by $Q\phi_n$. The change it induces is
\begin{equation}
\vev{\partial c c \partial X,Q \phi_n}=-\vev{Q(\partial c c \partial X),
     \phi_n}=0\,.
\end{equation}
This, however, raises a puzzle, as we are using a similar form for our initial
condition, $\psi_1=Q X(0)\ket{0}$. So it may seem that $A_1=0$, which is obviously
wrong. This stems from the fact that $X\ket{0}$ lies outside the Hilbert
space. The operator $x_0$ turns the usual infinite volume factor $\delta(p)$
into $\delta'(p)$. This works as follows, we add momentum dependence to the
bra state. Now,
\begin{align}
\nonumber
\vev{(\partial c c \partial X)(z_1) e^{i pX}(z_2),Q X}&=
  -\vev{(\partial c c \partial X)(z_1) Q(e^{i pX})(z_2),X}\\
&=-\vev{(\partial c c \partial X)(z_1) 
     \big(p^2\partial c e^{ipX}+ip c\partial X e^{ipX}\big)(z_2),X}.
\end{align}
Here, we use point split regularization due to the appearance of the
$\partial X$ term. The $\partial X \partial X$ singularity is exactly
canceled by the ghosts and we are left with the matter correlator
\begin{align}
\nonumber
2ip \vev{e^{ipX},X}=-2p\partial_q\vev{e^{ipX},
       e^{-iq X}}\big|_{q=0}=2p\delta'(p)=-2\delta(p)\,,
\end{align}
in agreement with~(\ref{A1}).
All the other coefficient calculations that we performed are
not affected by this subtlety, since there the zero mode cancels out.

\section{Other marginal deformations}
\label{OtherSolustion}

The solution that we constructed is for the marginal deformation related to
the $\partial X$ operator. However, we can generalize to other deformations
with the same OPE, such as the $\cos(X)$ deformation.
To realize how should the generalization look like we recall the logic of
our construction.
To find the $\partial X$ solution we introduced a primitive for
$\partial X$, including a zero mode $x_0$, such that $Q X=c\partial X$.
This enlargement of the Hilbert space allowed us to treat closed, non-exact
states as if they were exact.
Next we required that the solution does not depend on this zero mode, i.e.,
that it is well defined in the small Hilbert space.
To that end we had to add counter terms of the form of powers of the primitive.

All that works exactly the same for $\cos(X)$. We first define the primitive
\begin{align}
\label{Xi}
\Xi(z)=\Xi_0+\int_0^z \cos(X(\tilde z))d\tilde z\,.
\end{align}
The form of the solution is the same as before, and
the $\Xi_0^n$ terms drop out for same reasons.
Since the result is $\Xi_0$ independent, it is given just
by integrals of the $\cos(X)$ operator. Interestingly, in this representation,
which can also be applied to the $\partial X$ case, there are $n-1$ integrals
at order $n$, just like in the solutions~\cite{Schnabl:2007az,Kiermaier:2007ba}.
However, while there the integrals represent changing the size of a wedge state,
here the wedge state is fixed and the integrations are over the positions of the
insertions.

This analysis is, however, too naive. The problem is related to the issue of
normal ordering. While the OPE's $\partial X \partial X$ and
$\cos(X)\cos(X)$ have the same singular structure, they have different
finite contributions, which alters the form of the solution.
A possible resolution is to work with the $\cos(X)$ operator using a different
normal ordering scheme, related to the modes in the expansion of its
primitive. However, there are some subtleties in this route as well.
We believe that these subtleties are of a technical nature, especially in
light of the fact that the operators $\partial X$ and $\cos(X)$ are
related by an $SU(2)$ transformation and have the same form of commutation
relation with $Q$.

We would face similar problems in generalizing the construction
to deformations such as $e^{X_0}$, whose OPE is regular. Again,
it seems that defining a primitive and modifying the normal ordering
scheme should be enough for defining a solution. We hope to return
to these issues in the near future.

Our construction should work only for deformations which are
exactly marginal. At first instance it may seem that we could
generalize the construction also for vertex operators whose OPE
behaves like
\begin{equation}
VV\sim z^{-2}+z^{-1}V\,.
\end{equation}
The caveat here is that the structure functions associated with
normal ordering $V$ are $V$ dependent. Therefore, the proof
below~(\ref{NormalOrder}) regarding the regularity of the normal
ordered expression, does not generalize to this case.

The one case where our construction can be directly generalized is
$\partial X_+$. Inspecting~(\ref{Qexp}) for this case shows
that~(\ref{QX}) is replaced in this case by
\begin{equation}
[Q,X^n_+]=n c \partial X_+ X_+^{n-1}\,.
\end{equation}
The form of the solution is the same, except that there are no
$\partial c$ terms. Also, the terms
that arose from normal ordering are now absent. This leads to a
simplification in the evaluation of coefficients.
In particular, for the tachyon and the light-cone photon we get
\begin{equation}
A_n=0 \quad \forall n>1\,,\qquad
T_n=0 \quad \forall n\,.
\end{equation}
It would be interesting to compare this solution to the recently
found one~\cite{Erler:2007rh}.

\section{Comparing with former solutions}
\label{sec:previous}

In this section we comment on the relation of our construction
to that of Kiermaier, Okawa, Rastelli and Zwiebach
and of Schnabl~\cite{Kiermaier:2007ba,Schnabl:2007az}.
The construction of these authors starts
with any BRST-closed state in the Schnabl gauge
\begin{equation}
Q\psi_1=0\, \qquad \cB_0\psi_1=0\,.
\end{equation}
Alternatively, one could work in the Siegel gauge or any other gauge,
but the choice of Schnabl's gauge simplifies later calculations.
A one-parameter family of solutions can be defined through
\begin{equation}
\Psi = \sum_{n=1}^\infty \lambda^n\psi_n\,,
\end{equation}
with
\begin{equation}
\label{psiBL}
\psi_n \equiv -\frac{\cB_0}{\cL_0} \sum_{k=1}^{n-1}\psi_k\psi_{n-k}\,.
\end{equation}
This solution might be ill defined because the $1/\cL_0$ operation
is not a priori well defined,
as $\cL_0$ has a nontrivial kernel.
Consequently, $\cB_0/\cL_0$ is only well-defined up to an addition of
an arbitrary BRST-closed state.
In fact, in the construction of~\cite{Kiermaier:2007ba,Schnabl:2007az}
adding such a term in some cases becomes necessary in order
to cancel singularities.

Adding BRST-closed states at each order is equivalent
to adding them all at the first stage, provided
that an explicit $\lambda$ dependence of $\psi_n$ is allowed.
That is, given
\begin{equation}
\label{psiLambdaChi}
\tilde \psi_n \equiv -\frac{\cB_0}{\cL_0}
 \sum_{k=1}^{n-1}\tilde \psi_k \tilde \psi_{n-k}+
\chi_n\,,
\end{equation}
where $\chi_n$ are closed and $\tilde \psi_n$ are $\lambda$-independent,
we obtain a solution $\tilde\Psi$ which equals the state $\Psi$
built from $\psi_n(\lambda)$, where
\begin{equation}
\label{psiLambda}
\psi_{n>1} = -\frac{\cB_0}{\cL_0}
 \sum_{k=1}^{n-1}\psi_k \psi_{n-k}\,,\qquad
   \psi_1=\sum_{n=1}^\infty \lambda^{n-1} \chi_n\,.
\end{equation}
This should be clear, since it simply amounts to a $\la$ expansion.
A rigorous proof is given in Appendix~\ref{App:proof}.

The identity for $\psi_1$ in~(\ref{psiLambda}) expresses exactly
the way two solutions that are identical at lowest order (i.e., have
identical $\chi_1$) may differ at higher levels.
Namely, we are free to take $\chi_{n>1}$
equal to any BRST-closed state.
Adding to $\chi_n$ a term proportional to $\chi_1$
is equivalent to 
a redefinition of the deformation parameter as in~(\ref{CFTSFT}).
Adding to $\chi_n$ another BRST-closed state (corresponding to a linear
combination of other linearized marginal deformations)
leads, however, to a physically inequivalent solution.
Only the addition of a BRST-exact state leads
to a solution that is gauge equivalent to the original one.
It is useful to note at this point that fixing the gauge only removes
the ambiguity of adding a BRST-exact state, not of a more general,
BRST-closed one. However, the $\cB_0$ operator does not fix the gauge
completely. Indeed, a pure gauge state satisfying the $\cB_0$ gauge
is the basis of Schnabl's solution.

In light of these considerations it is interesting to try and
compare our solution to the solutions
of~\cite{Kiermaier:2007ba,Schnabl:2007az}.
The latter obey the Schnabl gauge by construction.
While this gauge choice breaks down
when counter terms are added, it is true for the regular solutions.
Our solutions on the other hand do not obey the $\cB_0$ gauge.
The general form of our
solution is $\hat U_{n+1}$ acting on a combination of $c$ and $\partial c$
insertions at various points acting on matter insertions.
The commutation relations
\begin{equation}
\cB_0 \hat U_n = \hat U_n
    \Big( \frac{4-n}{2}\cB_0 + \frac{2-n}{2}\cB_0^\dagger \Big) \,,\qquad
[\cB_0,c(z)]=z\,,
\end{equation}
imply that already $\phi_2$ does not obey the gauge condition
even before counter terms are added, and adding them cannot alter this result.

Consequently, our solution and those of~\cite{Kiermaier:2007ba,Schnabl:2007az}
are in different gauges.
Nevertheless, both solutions are built only by using the original
linearized marginal solution $\psi_1$ and string field products thereof,
and don't involve other linearized marginal solutions.
For this reason we expect that these solutions are
gauge equivalent, possibly after a $\la$ reparametrization.

\section{Conclusions}
\label{sec:conc}

We have presented the first explicit analytic solution corresponding to
the photon marginal deformation $\partial X$.
Our method leads, at first instance, to a non-normalizable dependence
of the solution on the center of mass coordinate. Key part of our
construction is that it is carefully tuned in order to cancel this dependence,
leading to a solution in the physical Hilbert space.
For a compact spatial direction, this deformation corresponds to giving
a vev to the Wilson loop. The center of mass coordinate $x_0$
should be regarded in this context as a formal extension of the
algebra.

One issue that remains open is the determination of the radius of
convergence of the series defining the solution.
Examination of the lowest-order coefficients for the tachyon state
and the photon state suggests that this radius of convergence is finite.
Another related issue is to find the functional relation between $\la_{CFT}$
and $\la_{SFT}$ used here.

While we presented explicitly the photon marginal deformation,
is should also be possible to use our method for other
operators with the same OPE, such as $\cos(X)$ or $\cosh(X^0)$.
It would be interesting to identify the
periodicity with respect to $\la_{CFT}$ in terms of $\la_{SFT}$.
Also, since this marginal
deformation connects the perturbative and the true vacua, it should
give a new representation of the true vacuum and it may even be
possible to explore the physics around it.
Our method is particularly advantageous for operators that have a
singular OPE in the coincidence limit, as it avoids corresponding
singularities in the solution.
However, it is also possible to use it for describing
marginal deformations with regular OPE.

An interesting extension of our method would be to try and
construct space-time dependent solutions, with a nontrivial but
normalizable dependence on the center of mass coordinate.
Another exciting possibility would be to use it in the context of
supersymmetric string field theory. This may prove to be quite natural
in light of the recent findings of~\cite{Erler:2007rh,Okawa:2007ri}.

\section*{Acknowledgments}

We benefited much from discussions with
Nathan Berkovits, Ted Erler, Stefan Fredenhagen, Sanny Itzhaki,
Yuji Okawa, Yaron Oz, Martin Schnabl, Ashoke Sen, Shigenori Seki
and Stefan Theisen.
E.~F.\ and M.~K.\ would like to thank the organizers of the HRI
string workshop and of the ISM06
for providing stimulating research environments in enjoyable atmosphere
and to Tel-Aviv University for hospitality.
R.~P.\ thanks the Albert-Einstein-Institute for hospitality and financial
support during his visits.
The work of M.~K.\ is supported by a Minerva fellowship.
The work of E.~F.\ is supported by the DIP foundation.
The work of R.~P.\ is supported by the Portuguese Funda\c c\~ao para a
Ci\^encia e a Tecnologia.

\appendix
\section{Proving the equivalence of~(\protect\ref{psiLambdaChi}) and~(\protect\ref{psiLambda})}
\label{App:proof}

To prove that adding BRST-closed states order by order
is equivalent to adding them all at the first stage, we expand
\begin{equation}
\psi_n = \sum_{k=0}^\infty \psi_{n,k} \la^k\,.
\end{equation}
By expanding~(\ref{psiLambda}) with respect to $\la$ we get that
these states obey the recursion relation
\begin{equation}
\label{psi_nkRec}
\psi_{n>1,k} = -\frac{\cB_0}{\cL_0}
 \sum_{j=1}^{n-1}\sum_{l=0}^{k}\psi_{j,l} \psi_{n-j,k-l}\,.
\end{equation}
We now define
\begin{equation}
\label{psi_nk}
\tilde \psi_n \equiv \sum_{m=1}^n \psi_{m,n-m}\,.
\end{equation}
What we have to prove is that these states obey the recursion
relation~(\ref{psiLambdaChi}).
For $n=1$ this hold trivially. For $n>1$, we
plug~(\ref{psi_nkRec}) into~(\ref{psi_nk}) and get
\begin{equation}
\tilde \psi_n = -\frac{\cB_0}{\cL_0}\left[
 \sum_{m=2}^n \sum_{j=1}^{m-1} \sum_{l=0}^{n-m}
  \psi_{j,l} \psi_{m-j,n-m-l}\right]+
\chi_n\,,
\end{equation}
where we separated the contribution from $m=1$.
Comparing to~(\ref{psiLambdaChi}), we see that what we have to prove is
that the expression in the square brackets equals
$\sum_{k=1}^{n-1}\tilde \psi_k \tilde \psi_{n-k}$.
Note that nowhere did we use the equation of motion or any property
of the operator in front of the square brackets other than being linear
and well defined. Neither did we use the closeness of the $\chi_n$'s.
Now write this expression as
\begin{equation}
 \sum_{j=1}^{n-1} \sum_{r=1}^{n-j} \sum_{l=0}^{n-j-r}
  \psi_{j,l} \psi_{r,n-j-r-l}=
 \sum_{j=1}^{n-1} \sum_{k=j}^{n-1} \sum_{r=1}^{n-k}
  \psi_{j,k-j} \psi_{r,n-k-r}=
 \sum_{k=1}^{n-1}\tilde \psi_k \tilde \psi_{n-k}\,,
\end{equation}
where we first interchanged the order of the first and second sums
and defined $r=m-j$. Then, we defined $k=l+j$ and interchanged the
second and third sums. Finally, we interchanged the order of the first
two sums and used the definition~(\ref{psi_nk}). This completes the proof.

\bibliography{FK}

\end{document}